\documentstyle{mn}
\input{epsf}

\title[The X-ray spectrum of 4U 1608-52] 
{The X-ray spectrum of the atoll source 4U 1608-52}

\author[M. Gierli\'nski, C. Done]
{Marek~Gierli\'nski$^{1,2}$ and Chris Done$^1$\\
$^1$Department of Physics, University of Durham, South Road, Durham DH1 3LE, 
UK\\ 
$^2$Obserwatorium Astronomiczne Uniwersytetu Jagiello{\'n}skiego, 30-244 
Krak{\'o}w, Orla 171, Poland}

\date{Submitted to MNRAS}
\pagerange{\pageref{firstpage}--\pageref{lastpage}}
\pubyear{2001}

\begin{document}

\topmargin = -0.5cm

\maketitle

\label{firstpage}

\begin{abstract}

The transient atoll source 4U 1608--52 had been extensively observed 
by the Rossi X-ray Timing Explorer ({\it RXTE\/}) during its 1998 
outburst. We analyse its X-ray spectra as a function of
inferred accretion rate from both the 1998 outburst and from 
the non-outburst 1996 and 1998 data. We can fit all the spectra by a 
model in which seed photons from the neutron star surface are 
Comptonized in a boundary layer. The Comptonized emission illuminates 
the accretion disc surface, producing an ionized, relativistically 
broadened reflection signature, while the direct emission from the 
accretion disc can also be seen.  The evolution of the source can be 
explained if the main parameter driving the spectral evolution is the 
average mass accretion rate, which determines the truncation radius 
of the inner accretion disc. At low mass accretion rates, in the 
island state, the disc truncates before reaching the neutron star 
surface and the inner accretion flow/boundary layer is mostly 
optically thin. The disc emission is at too low a temperature to be 
observed in the {\it RXTE\/} spectra, but some of the seed photons 
from the neutron star can be seen directly through the mostly 
optically thin boundary layer. At higher mass accretion rates, in the 
banana state, the disc moves in and the boundary layer becomes much 
more optically thick so its temperature drops. The disc can then be 
seen directly, but the seed photons from the neutron star surface 
cannot as they are buried beneath the increasingly optically thick 
boundary layer.

\end{abstract}



\section{Introduction}
\label{sec:introduction}

Low Mass X-ray Binaries (LMXBs) with a neutron star primary can be 
observationally divided into two main categories, namely atolls and Z 
sources (Hasinger \& van der Klis 1989).  This classification is 
based on changes in both spectral and timing properties as the source 
varies, and probably reflects differences in both mass accretion 
rate, $\dot{M}$, and magnetic field, $B$. The Z sources have high 
luminosity (typically more than 50 per cent of the Eddington limit) 
and magnetic field ($B > 10^9$ G) while the atolls have lower 
luminosity (generally less than 10 per cent of Eddington) and low 
magnetic field ($B \sim 10^8$ G) (Hasinger \& van der Klis 1989).

Both atolls and Z-sources show spectral changes which form a well 
defined track in a colour-colour diagram. The atolls show hard, 
power-law spectra at low luminosities (island state), which are 
startlingly like the low/hard spectra of Galactic black hole binaries 
(e.g. Barret \& Vedrenne 1993; van Paradijs \& van der Klis 1994; 
Barret et al.\ 2000). At higher luminosities the spectra are 
typically much softer (banana state; e.g.\ di Salvo et al.\ 2000; 
Piraino, Santangelo \& Kaaret 2000). The Z sources always show soft 
spectra, but there are subtle spectral changes which show up as a 
Z-shaped track on a colour-colour diagram. Both types of sources move 
{\em along\/} their tracks on the colour-colour diagram on time-scale 
of hours to days, except the island state of atolls, where motion can 
take days or weeks. They do not jump between the track branches. Most 
of the X-ray spectral and timing parameters depend only on the 
position of a source in this diagram. This is usually parameterized 
by the curve length, $S$, along the track. 

Until recently the atolls were thought to show a characteristic C- 
(or atoll) shaped track on the colour-colour diagram. However, 
mapping the full range of colour-colour changes for these LMXBs is 
difficult because the data available from most of the sources show 
little variability. There are only a few transient LMXBs with enough 
observations covering wide range of luminosities, to follow the whole 
track traced out in the colour-colour diagram. In these transient 
systems the C-shaped path is only a subset of a larger track. At low 
luminosities, evolution {\em within} the island state forms an upper 
branch, turning the C into a Z (Gierli{\'n}ski \& Done 2002; Muno, 
Remillard \& Chakrabarty 2002).

While the colour-colour diagram is clearly useful as a way of 
parameterizing the spectral changes, true spectral fitting gives much 
more information on the underlying physical changes in the source 
emission. Here we look at the detailed spectral shape of the 
transient atoll 4U 1608--52 as a function of its position on the 
colour-colour diagram, and show for the first time that {\em all\/} 
the spectra can be fit with a model in which a Comptonized boundary 
layer around the neutron star illuminates the accretion disc, giving 
rise to a relativistically smeared (and often highly ionized) 
reflected component and iron line. We show that the spectral changes 
mapped by the colour-colour diagram are consistent with a picture in 
which increasing average mass accretion rate increases the optical 
depth of the boundary layer and decreases the inner radius of the 
accretion disc.


\section{The source}
\label{sec:the_source}

4U 1608--52 is one of the transient neutron star LMXBs, with 
outbursts triggered by the disc instability which reoccur on time 
scales of $\sim 80$ days to 2 years (Lochner \& Roussel-Dupr{\'e} 
1994). However, little is known about the system as the optical 
counterpart is very faint (Grindlay \& Liller 1978; Wachter 1997). 
The distance is likewise poorly known. The current best estimate of 
3.6 kpc is from observations of flux saturated (radius expansion, 
Eddington limited) type I X-ray bursts (e.g. Nakamura et al. 1989). 
The inclination is constrained only by the fact that there are no 
eclipses in the X-ray light curve. The neutron star spin may be 
indicated by the recently detected burst oscillation in {\it RXTE\/} 
at 620 Hz (Muno at al.\ 2001)

Hasinger and van der Klis (1989) classified 4U 1608--52 as an atoll 
source from its spectral and timing properties. Mitsuda et al.\ 
(1989) observed with {\it Tenma\/} a continuous spectral transition 
from the banana high-luminosity, soft thermal spectrum to the island 
state with its low-intensity, hard power-law spectrum. They 
successfully fit these spectra by a two-component model consisting of 
the multicolour blackbody from the accretion disc and the Comptonized 
blackbody from the boundary layer. Higher energy observations with 
BATSE show that the island state power law is cut-off above $\sim$ 
100 keV (Zhang et al.\ 1996). An iron fluorescence line is detected 
in the {\it Tenma\/} X-ray spectra (Suzuki at al.\ 1984) with 
intensity weakly correlated with the source flux (Hirano et al.\ 
1987). {\it Ginga\/} observations showed spectra in varying stages 
between the island and banana states, where the iron line (modelled 
including the Compton reflected component) increased as the spectrum 
softened and the source moved towards the banana branch (Yoshida et 
al.\ 1993).

The power density spectrum of 4U 1608--52 in the island state 
resembles that of the black holes in the hard state (Yoshida et al.\ 
1993). It is characterized by a flat top noise ($P_\nu \propto 
\nu^{s}$, with the slope $s \sim 0$) below and a turn-over above a 
certain frequency, $\nu_b \sim$ 0.2--1 Hz. There is also a 
low-frequency quasi-periodic oscillation (QPO) around 2--9 Hz. {\it 
RXTE\/} has seen a similar spectrum, where the high-frequency noise 
can be represented by the exponentially cut-off power law (Yu et al.\ 
1997). In the banana state the power spectrum consists of very 
low-frequency noise below $\sim$ 1 Hz (power law with $s \sim -1$), 
and an exponentially cut-off power law at higher frequencies
(M{\'e}ndez et al.\  1999).

Twin kilohertz QPO's were detected by {\it RXTE\/} (van Paradijs et 
al.\ 1996; Berger at al.\ 1996; M{\'e}ndez et al.\ 1998a) with 
varying peak separation (M{\'e}ndez et al.\ 1998b). A third kilohertz 
QPO has also been detected (Jonker, M{\'e}ndez \& van der Klis 2000). 
The kHz QPOs are only found in a limited range of the colour-colour 
diagram, near the transition between the island and banana states 
(M{\'e}ndez et al.\ 1999). 


\section{Data analysis}
\label{sec:data_analysis}

We have analyzed {\it RXTE\/} observations from the 1998 
February--April outburst (archival numbers P30062, P30188, and 
P30502) and from low-luminosity observations in 1996 December 
(P10094) and 1998 June, August and September (P30419). All these data 
belong to the Proportional Counter Array (PCA) gain Epoch 3.  A 
conservative prescription to characterise uncertainties in the 
instrument response is to add 2 per cent systematics to the data for 
bright sources (e.g.\ Cui et al.\ 1997), irrespective of which 
detector/layer combination and gain epoch are used.  However, 
inspection of data from the Crab shows that in Epoch 3, when using 
{\em only\/} top layer data from detectors 0 and 1, smaller 
systematics of 0.5 per cent are more appropriate (Wilson \& Done 
2001). We have repeated this Crab analysis, and found that even using 
a single absorbed (fixed at $N_H=3.2\times 10^{21}$ cm$^{-2}$; 
Massaro et al.\ 2000) power law gives a reduced $\chi^2=1.1$ with 0.5 
per cent systematic errors. This clearly shows that there are no 
residuals in the response (e.g.\ features from the Xenon edge at 4.7 
keV, or from the rapid decrease in effective area at low energies) of 
the top layer of detectors 0 and 1 at this level at this time. This 
is {\em not\/} the case for all layers and all detectors, where the 
reduced $\chi^2 = 2.2$ with 0.5 per cent systematic errors and there 
is a clear feature in the residuals at low energies. Thus we use {\em 
only\/} data from Epoch 3, and from the top layer of detectors 0 and 
1, and apply 0.5 per cent systematic errors. However, we have checked 
that our conclusions are robust even to using 1 per cent systematics, 
although this gives unrealistically low $\chi^2$ values. 

\begin{figure}
\begin{center} 
\leavevmode 
\epsfxsize=8cm \epsfbox{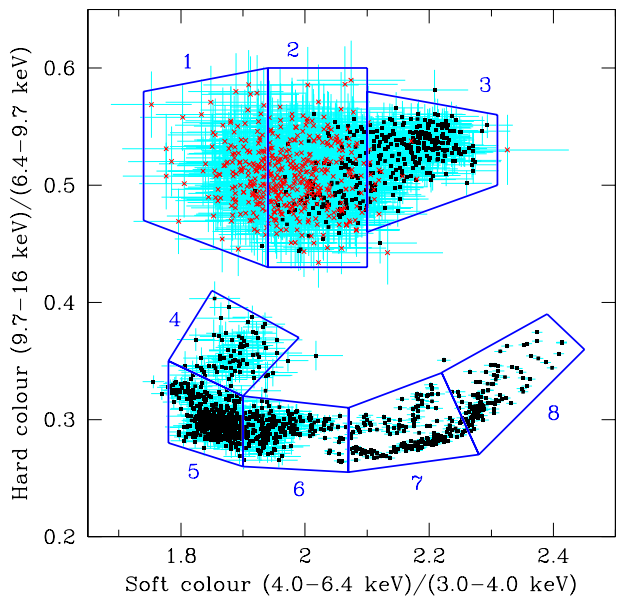} 
\end{center} 
\caption{Colour-colour diagram of 4U 1608--52. Each point represents 
128 s of the data. Slanted crosses correspond to the 
observations in 1996 December, when the source was at a rather 
low luminosity, while the squares correspond 
to the 1998 outburst data. Boxes show selection of the data for 
spectral analysis. Our choice of the boxes (and their order) follows 
a Z-shaped track in the colour-colour diagram, which is appropriate 
for the atolls, as it has been shown recently by Gierli{\'n}ski \& 
Done (2002).} 
\label{fig:colcol} 
\end{figure}

First, we created PCA light-curves for each energy channel in 128 s 
time bins. We excluded all the data affected by the type I bursts. 
This gave us 240 ks of PCA data. The light-curves were used to 
build a colour-colour diagram, defining a soft colour as a ratio of 
4--6.4 to 3--4 keV count rates, and a hard colour as a 9.7--16 over 
6.4--9.7 keV ratio. These count rates are corrected for the slow 
variations in the PCA response (van Straaten et al.\ 2000; see 
Gierli{\'n}ski \& Done 2002 for technical details applied here). The 
diagram is shown in Fig.\ \ref{fig:colcol}. 

Next, we extracted the PCA and High Energy X-ray Timing Experiment 
(HEXTE) spectra corresponding to different positions in the 
colour-colour diagram. We have assumed that 4U 1608--52 shows a 
Z-shaped track on the diagram, with inferred accretion rate 
increasing along the Z: from left to right on the upper branch, 
through the diagonal to the lower branch (Gierli{\'n}ski \& Done 
2002; Muno et al.\ 2002). The diagonal branch is sparsely covered in 
this data selection. We divided the diagram into eight regions 
(boxes), as shown in Fig.\ \ref{fig:colcol}, and averaged the spectra 
over each box. Neutron star LMXBs are known to vary in luminosity at 
given colours, creating the so-called parallel lines in 
colour-intensity diagrams (e.g.\ van der Klis 2001). Therefore, 
within one box we average spectra of different intensities. Though 
not necessary identical, these spectra should be very similar, 
considering nearness of colours. In the classic terminology of the 
atoll sources boxes 1--3 belong to the island state, boxes 5--8 to 
the banana, while the status of the box 4 cannot be established 
without joint spectral and timing analysis. To avoid accumulating 
spectral data from periods distant in time, we have created spectra 1 
and 2 from 1996 data only (slanted crosses in Fig.\ 
\ref{fig:colcol}), and spectra 3--8 from 1998 data only (filled 
squares in Fig.\ \ref{fig:colcol}). The PCA spectra are accompanied 
by the simultaneous HEXTE spectra extracted from cluster 0. We will 
refer to these joint spectra as S1--S8. 

For spectral fitting we use the X-ray spectral fitting package {\sc 
xspec} version 11 (Arnaud 1996). The error of each model parameter is 
given for a 90 per cent confidence interval. We use the 3--20 keV PCA 
data and 20--150 keV HEXTE data. The relative normalization of the 
PCA and HEXTE instruments is still uncertain, so we allow this to be 
an additional free parameter in all spectral fits. 


\section{Spectral model}
\label{sec:spectral_model}

Unlike black holes, the accreting neutron stars have both a surface 
and a magnetic field. Therefore, their X-ray spectra are expected to 
be more complex than these of the black holes. As well as emission 
from the accretion flow (direct emission from an optically thick, 
X-ray illuminated disc and/or an optically thin inner flow, corona or 
active regions; e.g.\ Czerny, Czerny \& Grindlay 1986; Esin, 
McClintock \& Narayan 1997; Beloborodov 1999) there can be emission 
from the neutron star surface and the boundary layer (e.g.\ Popham \& 
Sunyaev 2001). 

Observationally, the X-ray spectra of LMXBs in the soft state are 
well known to require at least two different spectral components, but 
the interpretation of these is not unique. The so-called {\em 
Eastern\/} model (Mitsuda et al.\ 1984, 1989) assumes that the soft 
component is from the disc, while the hard component is due to 
Comptonization in the boundary layer or the inner disc. In the {\em 
Western\/} model (White, Stella \& Parmar 1988) the soft component is 
a single-temperature blackbody from the surface, while the hard 
component is a Comptonized emission from the disc. While it is likely 
that both disc and boundary layer are Comptonized at some level, it 
seems more probable that the boundary layer should have a higher 
temperature than the accretion disc (Popham \& Sunyaev 2001). 

The hard state X-ray spectra of LMXBs are similar to those of the 
black hole candidates in their low/hard spectral state. They are 
dominated by a power law, with some contribution from an additional 
soft thermal component. The power law is usually interpreted as 
Comptonization of seed photons in hot, optically thin plasma (see 
e.g.\ Barret et al.\ 2000).

Thus, a physically motivated model which can potentially fit the hard 
component in {\it all\/} spectral states is thermal Comptonization. 
An optically thin, hot plasma would produce a power-law spectrum 
(island state), while a plasma with intermediate or high optical 
depth and lower temperature can create a softer spectrum in which the 
high energy rollover due to the electron temperature can be seen 
(banana state). We model the Comptonized spectrum using an 
approximate solution of Kompaneets (1956) equation (Zdziarski, 
Johnson \& Magdziarz 1996). The model (hereafter {\sc thcomp}, not 
included in the standard distribution of {\sc xspec}) is 
parameterized by the plasma electron temperature, $T_e$ and optical 
depth $\tau_e$ (or equivalently the asymptotic spectral index), and 
blackbody seed photon temperature, $T_{\rm seed}$. 

The seed photon energy is very important in the Comptonization model, 
since it gives a low-energy cut-off in the spectrum. For typical 
mass accretion rates onto a neutron star the seed photons from either 
the inner accretion disc or neutron star surface should have $kT_{\rm 
seed} \sim 1$ keV. This is plainly not far from the observed X-ray 
bandpass, so simple analytic approximations (e.g. the {\sc compst} 
model or a cut-off power law) for the Compton scattered spectrum are 
{\em not} valid, since they extend towards lower energies as a uncut 
power law (see e.g. Done, {\.Z}ycki \& Smith 2002). The model {\sc 
thcomp} we use in this paper treats the seed photons low-energy 
cut-off properly.

The boundary layer should illuminate the inner disc, producing a
relativistically broadened iron line and associated reflected
continuum. While a broad iron line is often identified in atoll
spectra, the reflected continuum which {\em must} accompany any line
emission has only been fit to island state spectra. We calculate the
Compton reflection of the {\sc thcomp} continuum spectrum using a
model where both the continuum reflection and the line are calculated
self-consistently for a given ionization state ({\.Z}ycki, Done \&
Smith 1998). The reflected spectrum (continuum and line) is also 
relativistically smeared for a given inner disc radius, $R_{\rm in}$. 
We approximate relativistic smearing by convolving the reflected 
spectrum with a relativistic line profile (Fabian et al.\ 1989).

\begin{figure*}
\begin{center} 
\leavevmode 
\epsfxsize=14cm \epsfbox{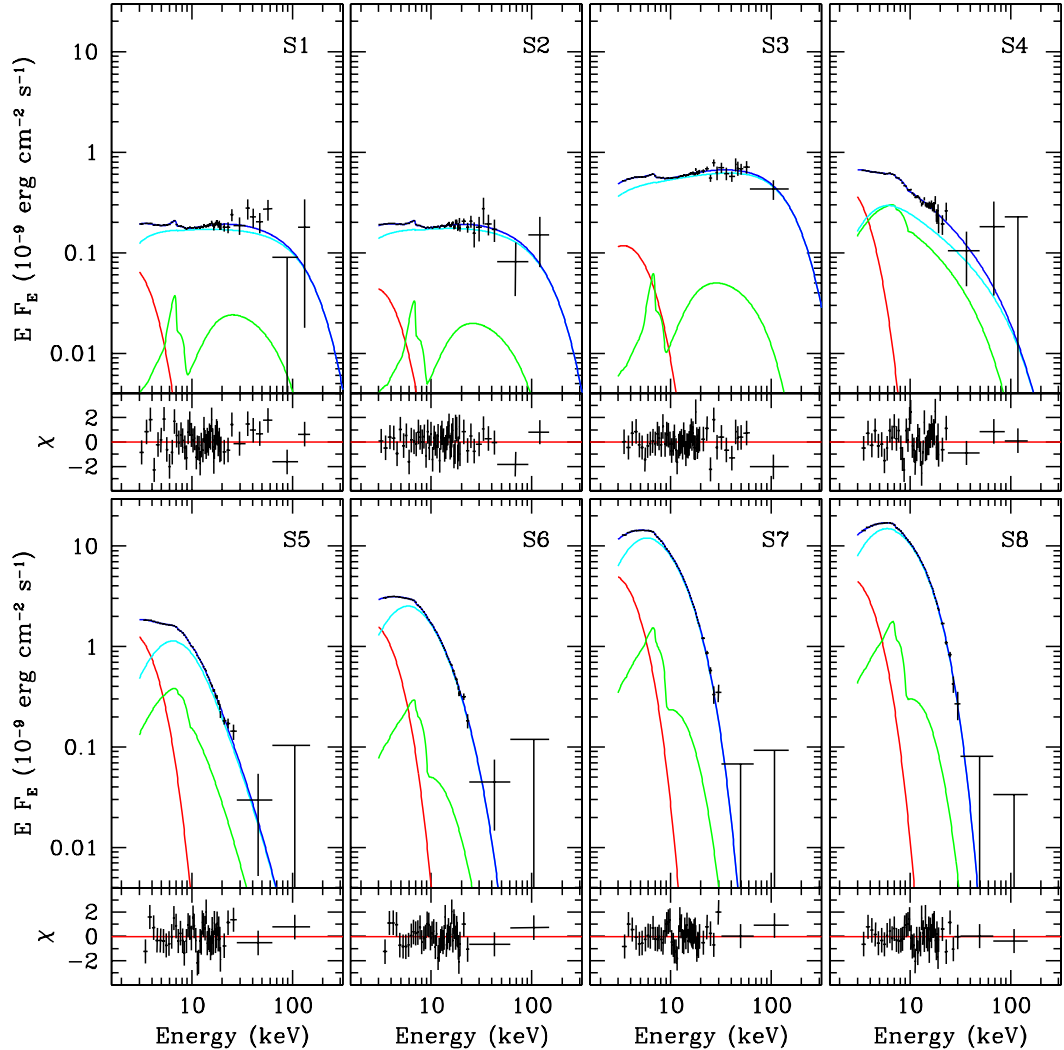} 
\end{center} 
\caption{PCA/HEXTE spectra of 4U 1608--52 along the atoll track, as 
selected from the data in Fig.\ \ref{fig:colcol}. The inferred 
accretion rate increases from S1 to S8. The individual panels show 
the unfolded data and models, as well as the residuals. The HEXTE 
data have been rebinned for clarity. The model, consisting of the 
disc blackbody (the soft component), Comptonization (the hard 
component) and its reflection, is described in Section 
\ref{sec:spectral_model}. The best-fitting parameters are shown in 
Fig.\ \ref{fig:pars}.}
\label{fig:spectra} 
\end{figure*}

First we demonstrate explicitly the need for a two component 
continuum in {\em all\/} the data sets. The {\sc thcomp} continuum 
model and its reflection alone cannot give satisfactory fits for the 
spectra S1--S5, where $\chi^2 > 120$ at 80 or 79 degrees of freedom. 
There is a strong soft excess in the residuals below $\sim$ 5 keV, 
which is particularly marked in S1 and S5 where $\chi^2 > 200$. We 
use a multicolour disc blackbody ({\sc diskbb} in {\sc xspec}; 
Mitsuda et al.\ 1984), to describe the soft component. This model is 
parameterized by the disc temperature, $T_{\rm soft}$, at the inner 
disc radius ($R_{\rm in}$) and normalization $N_{\rm soft} \propto 
R^2_{\rm in}$. However, its shape is poorly constrained by the {\it 
RXTE\/} data so it can equally well be a single-temperature 
blackbody. The spectra S6--S8 give good fits without the soft 
component ($\chi^2$ = 80, 88 and 75 at 78 d.o.f., for S6, S7 and S8, 
respectively), though even here inclusion of {\sc diskbb} improves 
the fits significantly. The soft component in the case of the weakest 
improvement (for S6 where $\Delta\chi^2 = 11.6$) is statistically 
significant at 99.7 per cent level.

The Compton reflection (including the self-consistent iron line 
emission) is always significantly detected also. Using the two 
component continuum model ({\sc diskbb+thcomp}) without a reflected 
component gives an increase in $\chi^2$ by more than 50 in all 
spectra (corresponding to an extremely high significance level).

These results motivate us to choose a model consisting of thermal 
Comptonization, its Compton reflection and the disc blackbody as the
model with which to fit {\em all\/} the spectra. 


\begin{table*}

\caption{Fitting results of the thermal Comptonization (with Compton 
reflection) plus disc blackbody model to the spectra S1--S8. The 
bolometric, unabsorbed flux, $F_{\rm bol}$, was estimated from the 
model. Units are as follows: $kT_{\rm soft}$ and $kT_{\rm seed}$ are 
in keV, $N_{\rm soft}^{1/2}$ is in km, $\xi$ is in erg cm s$^{-1}$ 
and $F_{\rm bol}$ is in $10^{-9}$ erg cm$^{-2}$ s$^{-1}$. Brackets in 
$kT_e$ column denote the fits in which $T_e$ was fixed, and its lower 
limits were calculated additionally (no upper limits were found). 
Results from this table are visualized in Fig.\ \ref{fig:pars}. A 
detailed description of the model used is in Section 
\ref{sec:spectral_model}. The individual spectra are selected from 
the data as shown in Fig.\ \ref{fig:colcol}. The spectra and the 
best-fitting models are shown in Fig.\ \ref{fig:spectra}.}

\label{tab:pars}
\begin{tabular}{cccccccccc}
\hline
Obs. & $kT_{\rm seed}$ & $\tau_e$ & $kT_e$ & $\Omega/2\pi$ & $\log(\xi)$ & $kT_{\rm soft}$ & $N^{1/2}_{\rm soft}$ & $F_{\rm bol}$ & $\chi^2$\\
\hline
1 & $0.51_{-0.51}^{+0.70}$ & $1.7_{-1.7}^{+0.9}$ & $(50)_{-22}$         & $0.16_{-0.06}^{+0.07}$ & $2.7_{-0.5}^{+0.5}$ & $0.68_{-0.06}^{+0.08}$ & $7.6_{-2.5}^{+2.7}$ & 1.0 & 81.7/78 \\
2 & $0.26_{-0.26}^{+1.06}$ & $1.7_{-1.7}^{+2.9}$ & $(50)_{-38}$         & $0.17_{-0.08}^{+0.09}$ & $2.5_{-0.7}^{+1.3}$ & $0.84_{-0.17}^{+0.06}$ & $3.5_{-0.5}^{+4.3}$ & 1.0 & 70.0/78 \\
3 & $0.26_{-0.26}^{+0.90}$ & $1.9_{-1.9}^{+1.3}$ & $(50)_{-26}$         & $0.07_{-0.03}^{+0.04}$ & $2.8_{-0.4}^{+0.8}$ & $1.16_{-0.16}^{+0.11}$ & $2.7_{-0.4}^{+1.3}$ & 3.1 & 75.0/77 \\
4 & $1.01_{-0.19}^{+0.23}$ & $1.0_{-1.0}^{+1.5}$ & $(50)_{-33}$         & $0.68_{-0.48}^{+1.32}$ & $4.9_{-0.8}^{+0.4}$ & $0.63_{-0.14}^{+0.13}$ & $22_{-8}^{+12}$     & 2.4 &101.9/77 \\
5 & $1.40_{-0.07}^{+0.07}$ & $0.4_{-0.4}^{+2.3}$ & $(50)_{-45}$         & $0.23_{-0.09}^{+0.12}$ & $4.6_{-0.3}^{+0.2}$ & $0.72_{-0.05}^{+0.05}$ & $28_{-4}^{+6}$      & 6.5 &106.1/77 \\
6 & $1.22_{-0.11}^{+0.14}$ & $3.6_{-2.5}^{+1.8}$ & $4.9_{-1.3}^{+11.5}$ & $0.08_{-0.02}^{+0.07}$ & $3.6_{-0.8}^{+0.9}$ & $0.75_{-0.09}^{+0.10}$ & $28_{-6}^{+12}$     & 9.7 & 68.2/76 \\
7 & $1.13_{-0.15}^{+0.25}$ & $6.0_{-1.7}^{+0.8}$ & $3.3_{-0.2}^{+0.7}$  & $0.09_{-0.02}^{+0.01}$ & $3.4_{-0.5}^{+0.6}$ & $0.80_{-0.11}^{+0.22}$ & $42_{-14}^{+43}$    & 36  & 52.8/76 \\
8 & $1.08_{-0.12}^{+0.21}$ & $7.2_{-1.1}^{+0.7}$ & $3.1_{-0.1}^{+0.3}$  & $0.09_{-0.02}^{+0.03}$ & $3.5_{-0.6}^{+0.7}$ & $0.74_{-0.20}^{+0.23}$ & $49_{-19}^{+60}$    & 42  & 55.3/76 \\
\hline
\end{tabular}
\end{table*}

\section{Results}
\label{sec:results}

For spectral fitting we use the model described in the previous 
section. We apply absorption corresponding to a fixed galactic 
hydrogen column of $N_H = 1.5\times10^{22}$ cm$^{-2}$ (Penninx et 
al.\ 1989). In the island state spectra, S1-S3, the HEXTE statistics 
at higher energies are not good enough to constrain the electron 
temperature of the Comptonized component, $T_e$. The high energy 
cutoff begins to be  seen in S4 and S5, although again the 
uncertainties on $T_e$ are large. All of these spectra provide only 
lower temperature limits. Therefore, we fit the spectra S1--S5 with 
$kT_e$ fixed at 50 keV, a value consistent with all these data sets, 
and then calculate the lower limits on the electron temperature. On 
the other hand, the spectra S6--S8 are able to constrain the electron 
temperature well. 

The data cannot constrain the inner disc radius from relativistic 
smearing of the reflected component. Spectral fitting gives only 
lower limits on $R_{\rm in}$, which are typically 10--20$R_g$ (where 
$R_g \equiv GM/c^2$) except for spectra S4 and S5 which are 
consistent with $R_{\rm in} = 6R_g$. We fix $R_{\rm in} = 20R_g$ in 
all the fits. 

\begin{figure}
\begin{center} 
\leavevmode 
\epsfxsize=7.5cm \epsfbox{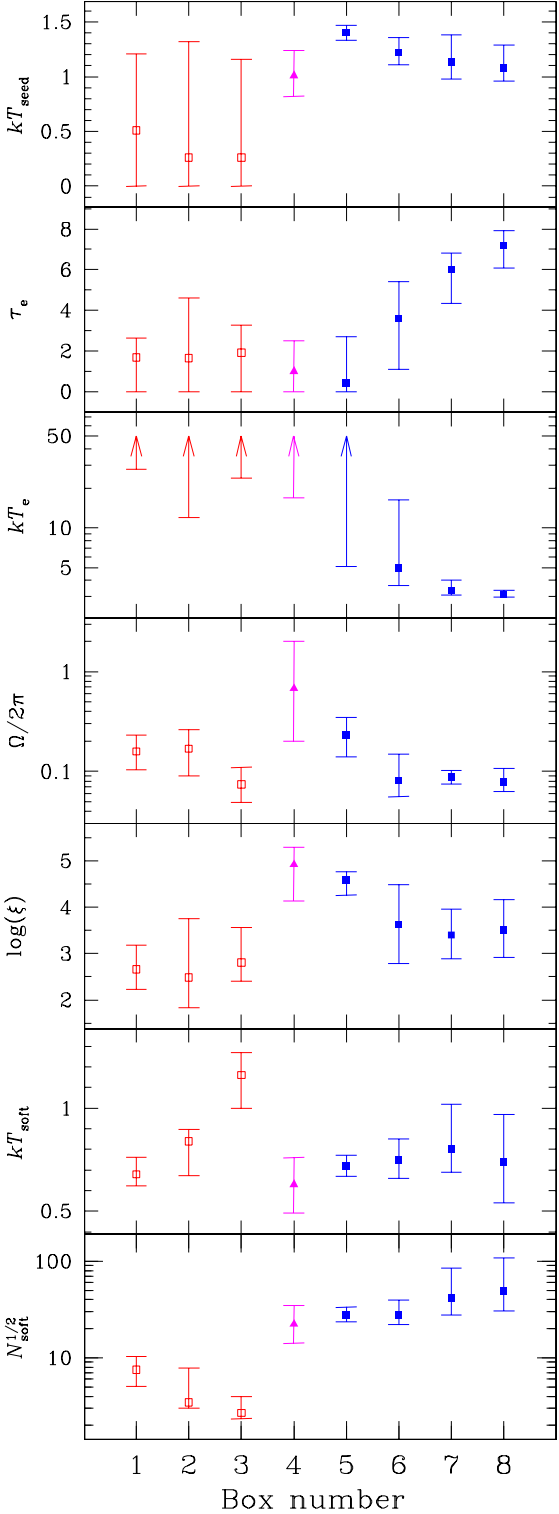} 
\end{center}

\caption{Results from Tab.\ \ref{tab:pars}. Units are as follows: 
$kT_{\rm soft}$ and $kT_{\rm seed}$ are in keV, $N_{\rm soft}^{1/2}$ 
is in km, $\xi$ is in erg cm s$^{-1}$ and $F_{\rm bol}$ is in 
$10^{-9}$ erg cm$^{-2}$ s$^{-1}$. While fitting spectra S1--S5 $T_e$ 
was fixed at 50 keV, and its lower limits were calculated 
additionally (no upper limits were found). The symbols used here 
correspond to the symbols in Figs.\ \ref{fig:tsoft_fsoft} and 
\ref{fig:tsoft_tseed}. We note that the box number on the horizontal 
axis does not exactly correspond to the curve length in the 
colour-colour diagram, and in particular, there is a gap between 
boxes 3 and 4.}

\label{fig:pars} 
\end{figure}

We fit all the spectra with the above constraints, and show the 
spectra unfolded with the best-fitting models in Fig.\ 
\ref{fig:spectra}. The best-fitting model parameters are shown in 
Tab.\ \ref{tab:pars} and Fig.\ \ref{fig:pars}. 

We calculate the unabsorbed bolometric flux, $F_{\rm bol}$, from the 
model. We cannot precisely estimate its uncertainties as this depends 
on the model used as well as on the statistics. However, fits with 
different soft component model (disc or blackbody) show consistently 
similar results, with the same trend: an increase from S1 to S3, a 
drop at S4 and a further increase from S5 to S8. This supports the 
inferred direction of the increasing accretion rate on the 
colour-colour diagram, tracing out a Z-shaped track (see also 
Gierli{\'n}ski \& Done 2002). 

With increasing inferred accretion rate the spectra become brighter,
softer, and have a decreasing electron temperature as seen from the
high energy cutoff. The change in the spectral shape between the island 
and banana states is clearly visible. Spectrum S4 bears more 
similarities to the banana state spectra than to island state, so it 
likely belongs to the banana state. The curvature expected from the 
low energy cut-off in the Comptonized spectrum close to the seed 
photon energy is clearly seen in spectra S4--S8, where $kT_{seed} > 1$
keV. The lower seed photon temperature derived for the island state 
means the low energy cutoff is less pronounced (not significantly
detected) in spectra S1-S3.

The electron temperature of the hard Comptonized component is 
consistent with a monotonic decline along the colour-colour track, 
while the optical depth is consistent with a corresponding monotonic 
increase. These two parameters are difficult to disentangle in data 
in which the high-energy rollover (at a few $kT_e$) is not observed, 
as in the island state spectra. However, while the optical depth 
derived for a given temperature also depends on the assumed geometry, 
it is clear that the island state spectra have $\tau_e$ of order 
unity, while the banana branch has $\tau_e \gg 1$. We derive similar 
optical depths and temperatures when we replace the approximate {\sc 
thcomp} thermal Comptonization model with {\sc compps} Comptonization 
code (Poutanen \& Svensson 1996), which finds a numerical solution of 
the Comptonization problem explicitly considering successive 
scattering orders.

Reflection (mostly driven by the detection of the self consistently 
produced iron line) is always significantly detected, although the 
inferred solid angle subtended by the reflecting material is 
generally rather low, with $\left<\Omega/2\pi\right> \approx 0.1$. 
The spectrum S4 has somewhat more reflection but this is not strongly
statistically significant. Reflection is always strongly ionized, 
with ionization increasing in the higher luminosity banana spectra. 
The presence of reflection in all spectra (both in island and banana 
states) demonstrates that the broad, ionized iron line seen from 
these sources can indeed be produced by illumination of the accretion 
disc. 

The rather small reflected fraction in the island state is in 
conflict with $\Omega/2\pi \approx$ 0.5--1 found in 2--20 keV {\it 
Ginga\/}  X-ray spectra of 4U 1608--52 by Yoshida et al.\ (1993) and 
Zdziarski, Lubi{\'n}ski \& Smith (1999). This is mostly due to the 
continuum model used. A power law and its reflected component does 
indeed give a larger $\Omega/2\pi$, but these fits are statistically 
unacceptable, as well as physically inconsistent. Comptonization {\em 
does not\/} give rise to a power law at energies close to the seed 
photon energies, and the data also require a soft component as well 
as the Comptonized emission. Our results are consistent (although 
with large uncertainties) with a reflection-spectral shape 
correlation for spectra S1-S4, as claimed by Zdziarski et al.\ (1999) 
although there is plainly no {\em overall\/} correlation that 
extends from the island state all the way through the banana branch.

\begin{figure}
\begin{center} 
\leavevmode 
\epsfxsize=7cm \epsfbox{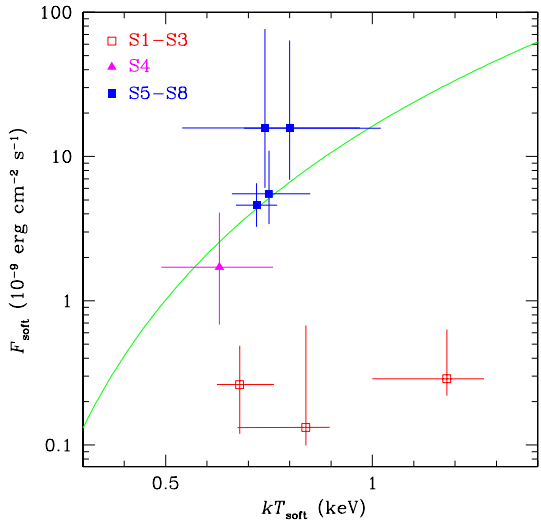} 
\end{center}
\caption{Soft component flux versus its temperature. The curve 
represents the best fit of a function $F_{\rm soft} = A T_{\rm 
soft}^4$ to the spectra S4--S8. The island state spectra S1--S3 are 
not consistent with this relation.}
\label{fig:tsoft_fsoft} 
\end{figure}

The observed soft component shows rather complex behaviour. Fig.\ 
\ref{fig:tsoft_fsoft} shows the soft component flux derived from the 
model given in Table 1, versus its temperature, $T_{\rm soft}$. There is a 
striking difference between the behaviour of the soft component in the 
island and banana states. The figure also shows the curve 
$F_{\rm soft} \propto T_{\rm soft}^4$, as expected if the disc geometry 
remained constant. The uncertainties are large, but while the banana 
state spectra could be consistent with this relation, the island state
is clearly not. This can be also seen from the fitting results (Fig.\ 
\ref{fig:pars}), where the normalization $N_{\rm soft} \propto R_{\rm 
in}^2$ is fairly constant along the banana track

\begin{figure}
\begin{center} 
\leavevmode 
\epsfxsize=7cm \epsfbox{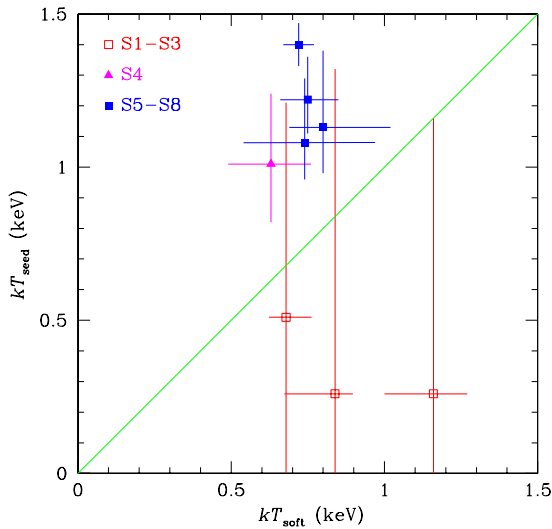} 
\end{center}
\caption{Seed photons temperature versus soft component temperature. 
The line shows were $T_{\rm seed} = T_{\rm soft}$. In the banana 
state (S4--S8) $T_{\rm seed}$ is systematically higher than 
$T_{\rm soft}$, therefore the soft component cannot be the source of 
seed photons. In the island state (S1--S3) the uncertainties are 
large, but it is possible that the soft component is the source of 
seed photons.}
\label{fig:tsoft_tseed}
\end{figure}

The difference in the soft component between the island and banana 
can be further seen in Fig.\ \ref{fig:tsoft_tseed}. The seed photons 
temperature is systematically higher then the soft component 
temperature in the banana state. Though 90 per cent confidence errors 
for $T_{\rm seed}$ and $T_{\rm soft}$ overlap (see Tab.\ 
\ref{tab:pars}), these two parameters are correlated and a fit with 
forced $T_{\rm seed} = T_{\rm soft}$ is significantly worse. In the 
island state the uncertainties on these temperatures are substantial 
but $T_{\rm seed}$ is consistent with being equal to $T_{\rm soft}$, 
and the soft component can be the source of seed photons. This result 
does not change when we use a single-temperature blackbody instead of 
the multicolour disc. Conversely, in the banana state, the observed 
soft component is {\em not\/} the source of seed photons for the 
Comptonized component.

In the island state, the observed seed photons (which are consistent 
with being the seed photons) have rather low normalization. If these 
were emitted from a disc then the inner disc radius can be estimated 
from the {\sc diskbb} model as \begin{equation} \label{eq:rin} R_{\rm 
in} \approx 0.61 \, N_{\rm soft}^{1/2} \,\, {2.7 \over \eta} {D \over 
{\rm 3.6~kpc}} \left({f_{\rm col} \over 1.8}\right)^2 \left({0.5 
\over \cos i}\right)^{1/2} {\rm km}, \end{equation} where $D$ is the 
distance to the source, $i$ is the inclination angle of the disc, 
$f_{\rm col}$ is the ratio of the colour to effective temperature 
(Shimura \& Takahara 1995) and $\eta$ is the correction factor for 
the inner torque-free boundary condition (Gierli{\'n}ski et al.\ 
1999: $\eta= 2.7$ for $R_{\rm in} = 6R_g$ and less for higher $R_{\rm 
in}$). With $\eta = 2.7$, $D$ = 3.6 kpc, $f_{\rm col} = 1.8$ and 
$\cos i = 0.5$ we find $R_{\rm in} < 5$ km for S1--S3. This is less 
than the expected neutron star radius of $\sim$ 10 km, and suggests 
that though we fit the soft component by a disc model, it might not 
be a disc at all. Even if the colour temperature correction is as 
high as $f_{\rm col} = 2.7$ (Merloni, Fabian \& Ross 2000), then S3 
still yields $R_{\rm in} < 5.5$ km. There are of course more unknowns 
in Eq.~(\ref{eq:rin}), in particular the inclination angle, but 
$R_{\rm in} > 10$ km requires $i > 81^\circ$, which is excluded by 
the lack of X-ray eclipses. 

It is therefore unlikely that the soft component is the accretion 
disc in the island state. Instead, we suggest it comes from the 
neutron star surface. A fit of a model including a single-temperature 
blackbody instead of a disc to the island state data yields an 
apparent radius of the blackbody of $4\pm2$ km. 

Conversely, in the banana states, the soft component {\em is\/} 
consistent with emission from the accretion disc, although these 
photons do {\em not} provide the seed photons for Compton scattering. 
The inner disc radius is consistent with remaining constant 
throughout the banana state. From the fit to the typical banana 
spectrum S6 we estimate $R_{\rm in} \approx 30$ km, a value 
consistent with the disc truncated not very far from the neutron 
star. 

If the soft component is the neutron star surface in the spectra 
S1--S3 and the accretion disc in the spectra S4--S8, then the change of 
trend in $T_{\rm soft}$ and $N_{\rm soft}$ between S3 and S4 (see 
Fig.\ \ref{fig:pars}) is naturally explained, and does not 
contradict our assumption about the direction of increase in the 
accretion rate in the upper branch in the colour-colour diagram.
We discuss a physical picture for the evolution of the soft and seed
photons in Section 7.

\section{Comparison to other sources} 
\label{sec:comparison}

The X-ray spectra of 4U 1608--52 analyzed here are entirely typical 
of atoll sources. Having a good coverage of a large luminosity 
variation, we can compare our results to previous observations of 
other atolls in various luminosity states.

The island state is characterized by a hard, power-law spectrum, 
similar to the hard state of black hole binaries, though a bit 
softer. The presence of a soft component has been reported before in 
SAX J1808.4--3658 (Gierli{\'n}ski, Done \& Barret 2002), 4U 1724--308 
(Guainazzi et al.\ 1998), 1E 1724--3045 and SLX 1735--269 (Barret et 
al.\ 2000). In these sources, the radius of the soft component is not 
consistent with the accretion disc (except perhaps SLX 1735--269, 
though there is Galactic bulge emission component in the spectrum, 
which makes spectral modelling difficult; see Barret et al.\ 2000). 
This is consistent with what we find in 4U 1608--52 and suggestive 
that the soft component arises from the neutron star surface rather 
than from the accretion disc.

The banana state X-ray spectra are soft, and here again the 4U 
1608--52 is entirely similar to other atolls. A two-component model 
has been successfully applied to e.g.\ KS 1731--260 (Narita, Grindlay 
\& Barret 2001), GX 3+1, Ser X-1 (Oosterbroek et al.\ 2001) and MXB 
1728--34 (di Salvo et al.\ 2000), the soft component being consistent 
with the disc. On the other hand, one must be very cautious here, 
since the soft component properties strongly depend on the 
Comptonization model applied, and inappropriate approximations have 
often been used (see Done et al.\ 2001 for details). 

The ambiguity in the spectral shape of the soft component has been 
reported before (e.g.\ Guainazzi et al.\ 1998; Barret et al.\ 2000). 
Interestingly, in MXB 1728--34, even {\it BeppoSAX\/} spectra which 
extend down to much lower energies than the PCA cannot distinguish 
between a disc and blackbody spectrum for the soft component (see di 
Salvo et al.\ 2000).

Compton reflection has been reported before in the island state, in 
e.g.\ SAX 1808.4--3658 (Gierli{\'n}ski et al.\ 2002), 4U 1728--34 
(Narita et al.\ 2001), 4U 0614+091 (Piraino et al.\ 1999), GS 
1826--238 and SLX 1735--269 (Barret et al.\ 2000). However, this 
paper shows the first detection of reflection in the banana state. 
Previous work has generally fit an {\it ad hoc} broad Fe K$\alpha$ 
line, rather than including the self consistently produced reflected 
continuum e.g.\ in GX 3+1, Ser X-1 (Oosterbroek et al.\ 2001) and 4U 
1728--34 (Piraino, Santangelo \& Kaaret 2000; Di Salvo et al.\ 2000).


\section{Discussion and conclusions}
\label{sec:discussion}

We fit the spectra of the atoll 4U 1608-52 at all points along its 
track on the colour-colour diagram by a physically motivated model, 
consisting of thermal Comptonization and its Compton reflection, 
together with a soft component. Both reflection (with the self 
consistently calculated iron line emission) and the soft component 
are always statistically significantly detected in the spectra.

There is a dramatic change in spectral shape between the island and 
banana states, probably caused by an equally dramatic change in the 
accretion flow geometry. In the island state the observed soft 
component luminosity and temperature implies an emission region which 
is smaller than an accretion disc around a neutron star.  We suggest 
it arises from optically thick, thermal emission from surface of the 
neutron star.  The hard component is consistent with Comptonization 
of these observed soft photons from the neutron star surface by hot 
electrons in an inner optically thin accretion flow (or outer 
boundary layer). The optically thin Comptonizing medium has 
temperature of $\ga$ 20 keV. We do not see direct emission from the 
accretion disc in the PCA energy range, but we do see weak reflected 
emission implying that the disc subtends a solid angle $\sim 0.1 
\times 2\pi$ from the point of view of the hard X-ray source.

In the banana branch spectra the Comptonized component is much softer,
with temperature of $\sim$ 5 keV and optical depth of $\gg 1$. Here
the observed soft component {\em is\/} consistent with being from the
disc, but the seed photons are {\em not} consistent with being the
observed soft component, and again it is more likely that the seed
photons are produced predominantly by the neutron star surface. This 
interpretation favours the Eastern model of LMXBs on the banana branch.

A scenario consistent with all of the above results is one in which 
the main parameter driving the spectral evolution is the average mass 
accretion rate, which determines the truncation radius of the inner 
accretion disc. At low mass accretion rates the optically thick disc 
does not extend down to the neutron star surface, and the inner 
accretion flow/boundary layer is mostly optically thin. The large 
radius of the optically thick disc means that its temperature and 
luminosity is low so it cannot be seen directly in the PCA spectra, 
and it subtends a rather small solid angle for reflection.  Since the 
accretion flow is not very optically thick then some fraction $\sim 
e^{-\tau}\sim 0.3$ of the emission from the neutron star surface can 
be seen directly, while the rest is Comptonized into the hard 
component. As the mass accretion rate increases, so the disc extends 
further in and the optical depth of the inner flow/boundary layer 
increases.  Eventually the neutron star surface is shrouded by the 
increasingly optically thick flow, so that these seed photons cannot 
be seen. The high optical depth also means that the boundary layer 
emission is then close to thermalizing, so its temperature drops, 
while the decreasing inner radius of the disc means that its emission 
starts to become more important in the PCA bandpass. 

In this scenario the island state/banana branch transition is marked
by the point at which the inner accretion flow collapses into a
disc. The disc extends further in so that it is observed more easily
as a soft component, softening the soft colour. But it also means that
the boundary layer becomes much more optically thick, so its
temperature drops, softening the hard colour.

It is plain that this can explain {\em qualitatively} the atoll
source evolution. We will investigate the {\em quantitative}
implications of this in a later paper. 

\section*{Acknowledgements}

This research has been supported in part by the Polish KBN grant
2P03D00514.


\label{lastpage}


\begin{thebibliography}{}


\bibitem[]{arn96} Arnaud K. A., 1996, in Jacoby G. H., Barnes J., eds.,
Astronomical Data Analysis Software and Systems V. ASP Conf. Series Vol.\ 101,
San Francisco, p.\ 17

\bibitem[]{bv94} Barret D., Vedrenne G., 1994, ApJSS, 92, 505

\bibitem[]{b00} Barret D., Olive J. F., Boirin L., Done C., Skinner 
G. K., Grindlay J. E., 2000, ApJ, 533, 329

\bibitem[]{bel99} Beloborodov A. M., 1999, ApJ, 510, L123

\bibitem[]{ber96} Berger M., et al., 1996, ApJ, 469, L13

\bibitem[]{cui97} Cui W., Heindl W. A., Rothschild R. E.,
 Zhang S. N., Jahoda K., Focke W., 1997, ApJ, 474, L57

\bibitem[]{ccg86} Czerny B., Czerny M., Grindlay J. E., 1986, ApJ, 
311, 241

\bibitem[]{dis00} Di Salvo T., Iaria R., Burderi L., Robba N. R., 
2000, ApJ, 542, 1034

\bibitem[]{dzs01} Done C., {\.Z}ycki P. T., Smith D. A., 2002, 
MNRAS, 331, 453

\bibitem[]{emn97} Esin A. A., McClintock J. E., Narayan R., 1997, 
ApJ, 489, 865

\bibitem[]{fab89} Fabian A., Rees M. J., Stella L., White N. E., 
1989, MNRAS, 238, 729



\bibitem[]{gd01} Gierli{\'n}ski M., Done C., 2002, MNRAS, 331, L47

\bibitem[]{g99} Gierli{\'n}ski M., Zdziarski A. A., Poutanen J., 
Coppi P. S., Ebisawa K., Johnson W. N., 1999, MNRAS, 309, 496

\bibitem[]{gdb01} Gierli{\'n}ski M., Done C., Barret D., 2002, 
MNRAS, 331, 141

\bibitem[]{gl78} Grindlay J. E., Liller W., 1978, L127

\bibitem[]{gua98} Guainazzi M., Parmar A. N., Segreto A., Stella 
L., dal Fiume D., Oosterbroek T., 1998, A\&A, 339, 802

\bibitem[]{hk89} Hasinger G., van der Klis M., 1989, A\&A, 225, 79

\bibitem[]{hir87}  Hirano T., Hayakawa S., Nagase F., Masai K., 
Mitsuda K., 1987, PASJ, 39, 619

\bibitem[]{jmk00} Jonker P. G., M{\'e}ndez M., van der Klis M., 2000, 
ApJ, 540, L29 

\bibitem[]{kkb96} King A. R., Kolb U., Burderi L., 1996, ApJ, 464, 
L127

\bibitem[]{kom56} Kompaneets A. S., 1956, Soviet Phys., JETP 31, 876

\bibitem[]{lr94} Lochner J. C., Roussel-Dupr{\'e} D., ApJ, 435, 840


\bibitem[]{mas00} Massaro E., Cusumano G., Litterio M., Mineo T.,
2000, A\&A, 361, 695

\bibitem[]{men98a} M{\'e}ndez M., et al., 1998a, ApJ, 494, L65

\bibitem[]{men98b} M{\'e}ndez M., van der Klis M., Wijnands R., Ford E. C., van 
Paradijs J., Vaughan B. A., 1998b, ApJ, 505, L23

\bibitem[]{men99} M{\'e}ndez M., van der Klis M., Ford E. C., Wijnands R., van 
Paradijs J., 1999, ApJ, 511, L49

\bibitem[]{mfr00} Merloni A., Fabian A. C., Ross R. R., 2000, MNRAS, 
313, 193

\bibitem[]{mit84} Mitsuda K., et al., 1984, PASJ, 36, 741

\bibitem[]{mit89} Mitsuda K., Inoue H., Nakamura N., Tanaka Y., 1989, PASJ, 41, 
97

\bibitem[]{mun01b} Muno M. P., Chakrabarty D., Galloway D. K., Savov 
P., 2001, ApJ, 553, L157

\bibitem[]{mun01a} Muno M. P., Remillard R. A., Chakrabarty D., 2002, 
ApJ, 568, L35

\bibitem[]{nak89} Nakamura N., Dotani T., Inoue H., Mitsuda K., 
Tanaka Y., Matsuoka M., 1989, PASJ, 41, 617

\bibitem[]{n01}  Narita T., Grindlay J. E., Barret D., 2001, ApJ, 
547, 420

\bibitem[]{ost01} Oosterbroek T., Barret D., Guainazzi M., Ford 
E. C., 2001, A\&A, 366, 138

\bibitem[]{pir99} Piraino S., Santangelo A., Ford E. C., Kaaret P., 
1999, A\&A, 349, L77 

\bibitem[]{psk00} Piraino S., Santangelo A., Kaaret P., 2000, A\&A, 
360, L35

\bibitem[]{ps01} Popham R., Sunyaev R., 2001, 547, 355

\bibitem[]{ps96} Poutanen J., Svensson R., 1996, ApJ, 470, 249

\bibitem[]{st95} Shimura T., Takahara F., 1995, ApJ, 445, 780


\bibitem[]{suz84} Suzuki K., Matsuoka M., Inoue H., Mitsuda K., 
Tanaka Y., Ohashi T., Hirano T., Miyamoto S., 1984, PASJ, 36, 761


\bibitem[]{kli01} van der Klis M., 2001, ApJ, 561, 943

\bibitem[]{pk94} van Paradijs J., van der Klis M., 1994, A\&A, 281, 
L17

\bibitem[]{par96} van Paradijs J., et al., 1996, IAUC 6336

\bibitem[]{vs00} van Straaten S., Ford E. C., van der Klis M., 
M{\'e}ndez M., Kaaret P., 2000, ApJ, 540, 1049

\bibitem[]{w97} Wachter S., 1997, ApJ, 485, 839 

\bibitem[]{w00} Wachter S., 1997, ApJ, 485, 839

\bibitem[]{whi88} White N. E., Stella L., Parmar A. N., 1988, ApJ, 
324, 363

\bibitem[]{wd00} Wilson C. D., Done C., 2001, MNRAS, 325, 167

\bibitem[]{Yos93} Yoshida K., Mitsuda K., Ebisawa K., Ueda Y., Fujimoto R., 
Taqoob T., Done C., 1993, PASJ, 45, 605

\bibitem[]{yu97} Yu W., et al., 1997, ApJ, 490, L153

\bibitem[]{zjm96} Zdziarski A. A., Johnson W. N., Magdziarz P., 1996, 
MNRAS, 283, 193

\bibitem[]{zls99} Zdziarski A. A., Lubi{\'n}ski P., Smith D. A., 
1999, MNRAS, 303, L11

\bibitem[]{zha96} Zhang S. N. et al., 1996, A\&AS, 120, 279

\bibitem[]{zds98} {\.Z}ycki P., Done C., Smith D. A., 1998, ApJ, 496, 
L25

\end{thebibliography}
\end{document}